\documentclass[11pt,a4paper]{article}

\usepackage[T1]{fontenc}
\usepackage[utf8]{inputenc}
\usepackage{lmodern}
\usepackage[a4paper,margin=2.55cm]{geometry}
\usepackage{amsmath,amssymb,amsfonts,mathtools,bm}
\usepackage{graphicx}
\usepackage{microtype}
\usepackage{booktabs}
\usepackage{caption}
\usepackage{float}
\usepackage{placeins}
\usepackage{enumitem}
\usepackage{xcolor}
\usepackage{hyperref}
\definecolor{linkblue}{RGB}{0,72,180}
\hypersetup{colorlinks=true,linkcolor=linkblue,citecolor=linkblue,urlcolor=linkblue,
  pdftitle={Foresight-dependent entropy production in stochastic two-agent dynamics},
  pdfauthor={Arpan Dey}}
\newcommand{\citelink}[1]{\textcolor{linkblue}{\cite{#1}}}
\newcommand{\figref}[1]{\hyperref[#1]{\textcolor{linkblue}{Fig.~\ref*{#1}}}}
\newcommand{\figrefs}[2]{\figref{#1} and \figref{#2}}
\newcommand{\reflink}[1]{\hyperref[#1]{\textcolor{linkblue}{Sec.~\ref*{#1}}}}
\newcommand{\applink}[1]{\hyperref[#1]{\textcolor{linkblue}{Appendix~\ref*{#1}}}}
\newcommand{\doilink}[1]{\href{https://doi.org/#1}{\textcolor{linkblue}{doi:#1}}}
\newcommand{\arxivlink}[1]{\href{https://arxiv.org/abs/#1}{\textcolor{linkblue}{arXiv:#1}}}
\newcommand{\tr}{\operatorname{tr}}
\newcommand{\E}{\mathbb E}

\newcommand{\ep}{\widehat{\sigma}}
\newcommand{\Delt}{\Delta B}
\newcommand{\Iflow}{\dot{\mathcal I}}
\allowdisplaybreaks[2]
\title{\bfseries Foresight-dependent entropy production in stochastic two-agent dynamics}
\author{Arpan Dey\\[3pt]\small Universit\'e de Montpellier, Montpellier, France\\[3pt]\small Email: \href{mailto:arpand2004@gmail.com}{\textcolor{linkblue}{arpand2004@gmail.com}}}
\date{}
\begin{document}
\maketitle
\begin{abstract}
Nonreciprocal interactions can break detailed balance by sustaining stationary probability currents. We ask whether such currents can also emerge when agents' immediate coordination preferences are reciprocal but their adjustment rules result from forward-looking optimization. We study a minimal stochastic linear--quadratic game in which two agents control noisy scalar states and minimize individual discounted costs in a full-information feedback Nash equilibrium. The optimized controls yield an Ornstein--Uhlenbeck process whose stationary current, entropy production, and inter-agent information flow follow exactly from the Nash feedback. For reciprocal coordination with unequal self-anchoring, the drift asymmetry first appears at order $\tau^5$, where $\tau$ is the planning timescale, and the normalized entropy production scales as $\tau^8$. We trace this delay to one mechanism: the asymmetry is set by the mismatch between the agents' value-function curvatures along the coordinate each does not control. At leading order this equals the mismatch of the corresponding cost curvatures, which instantaneous incentives cannot see and which the coordination cost fixes to zero; unequal persistence of the partners' fluctuations generates a value-function mismatch at order $\tau^3$, postponing the asymmetry to fifth order, while a direct stake in the partner's state restores a cubic law. The normalized information flow, of order $\tau^4$, identifies the weakly anchored agent as the one that tracks its partner. For conformist--contrarian coupling, normalized entropy production starts from a nonzero limit and initially decreases as $\tau^2$; for equal anchoring, we prove that its long-foresight limit lies below the short-foresight value and that the stationary mutual-information flow vanishes. Numerical solutions of the full Nash equations confirm these laws.
\end{abstract}

\section{Introduction}\label{sec:intro}
In stochastic systems, nonreciprocal interactions can sustain stationary probability currents and break detailed balance even when the probability distribution is time independent. Such behavior has been studied in linear Langevin models and interacting-particle systems \citelink{loos2020,zhang2023,fruchart2021,fruchart2026}. For multivariate Ornstein--Uhlenbeck (OU) processes, the stationary covariance, probability current, and entropy-production rate are known exactly in terms of the drift and diffusion matrices \citelink{godreche2019}. In those calculations the drift matrix is specified beforehand. Our question instead concerns the drift matrix \emph{produced} when each agent optimizes its own future costs while anticipating the other's response.

Strategic interaction and statistical mechanics have long been connected, for example through stochastic strategy-revision processes \citelink{blume1993}. Potential games admit a common function reproducing the agents' unilateral instantaneous incentives \citelink{monderer1996}. However, instantaneous potential structure does not generally imply that a full-state-feedback \emph{dynamic} game has a potential functional. Related distinctions between instantaneous incentives and dynamic potential-game structure have been examined for finite-horizon, discrete-time linear--quadratic (LQ) games \citelink{hosseinirad2026}; approximate potential structures in dynamic stochastic games have also been characterized \citelink{guo2024}. Coupled Riccati equations are well established in the study of feedback Nash equilibria of stochastic LQ differential games \citelink{sunyong2019}, including infinite-horizon formulations characterized by coupled algebraic Riccati equations \citelink{lishiyong2021}. In statistical physics, games in which each player controls Langevin dynamics with a cost quadratic in the control have been studied as quadratic mean-field games \citelink{swiecicki2016,ullmo2019}.

Related studies have used Nash solutions when individual entropy-production contributions enter the agents' objectives \citelink{fujimoto2024}, and have derived nonreciprocal continuum interactions from decision-making rules \citelink{lama2025}. The description of dynamics through gradients of individual ``selfish energies'' provides another connection between nonreciprocal response and distinct local objectives \citelink{loosjack2026}. A recent Hodge decomposition of discrete priority-coordination games links non-potential strategic components to stationary currents and entropy production under log-linear learning \citelink{lin2026}. In the present model, entropy production is instead an observable calculated \emph{after} finding the feedback Nash equilibrium, rather than a quantity that the agents optimize.

Our question is quantitative: \emph{given an exactly specified two-agent stochastic LQ game, how does the discount-controlled planning timescale change the stationary probability current of its feedback Nash dynamics, and why?} We examine two coupling classes within one framework: \emph{reciprocal coordination}, where each agent's instantaneous interaction cost favors matching the other, and \emph{conformist--contrarian coupling}, where one agent favors matching and the other favors opposition. In reciprocal coordination, the instantaneous costs admit a common potential and the Nash drift is reciprocal at leading short-foresight order; the linear and cubic contributions to its asymmetry cancel, and for unequal self-anchoring strengths the first nonzero term is fifth order. We show that this delay has a transparent origin. Foresight makes each agent's response sensitive to how much its partner's state costs it, a quantity to which instantaneous incentives are blind. The coordination cost equalizes these sensitivities, and only the unequal persistence of the partners' fluctuations unbalances them. The resulting current has a definite direction, which we quantify by the information flow between the agents \citelink{horowitz2014,loos2020}. In the conformist--contrarian class, the drift is nonreciprocal already at leading order, and foresight instead reduces normalized irreversibility.

\section{Stochastic model and feedback Nash equilibrium}\label{sec:model}
\subsection{Dynamics, costs, and assumptions}
Two agents, indexed by $i\in\{1,2\}$ with $j\ne i$, control scalar states $x_i\in\mathbb R$. Their stochastic dynamics are
\begin{equation}
 dx_i=u_i\,dt+\sqrt{2D}\,dW_i,\qquad
 \E[dW_i\,dW_j]=\delta_{ij}\,dt,
 \label{eq:sde}
\end{equation}
where $D>0$ is a common diffusion coefficient and $W_i$ are independent Wiener processes. The control $u_i$ sets the systematic rate of change of $x_i$, while the Wiener noise produces random fluctuations. Each agent minimizes its own running cost
\begin{equation}
 L_i(\bm x,u_i)=\frac a2(x_i-s_i x_j)^2
 +\frac{b_i}{2}(x_i-m_i)^2+\frac r2u_i^2,
 \qquad s_i\in\{+1,-1\},
 \label{eq:running}
\end{equation}
with $a,b_i,r>0$. The common $a$ fixes the interaction-cost strength; $b_i$ is the self-anchoring strength toward agent $i$'s own fixed preferred state $m_i$, and $r$ penalizes large adjustment rates. For $s_i=+1$, agent $i$ favors matching $x_j$; for $s_i=-1$, it favors the opposite state. We call $(s_1,s_2)=(+1,+1)$ \emph{reciprocal coordination} and $(+1,-1)$ \emph{conformist--contrarian coupling}. Neither label refers to a restriction on the actual control $u_i$, which follows from optimization.

The discounted infinite-horizon objective is
\begin{equation}
 J_i(\bm x_0;u_1,u_2)=\E_{\bm x_0}
 \int_0^\infty e^{-\rho t}L_i(\bm x(t),u_i(t))\,dt,
 \qquad \tau\equiv\rho^{-1},\quad \rho>0.
 \label{eq:objective}
\end{equation}
Here $\rho$ is the discount rate and $\tau$ the corresponding planning timescale. We seek time-independent full-state feedback Nash equilibria: each agent observes the current joint state $\bm x=(x_1,x_2)^T$ and chooses a control depending on that state alone. Both agents know the costs and equilibrium feedback rules and optimize against the other's rule. The resulting dynamics are Markovian, although future state realizations remain uncertain because of the noise in Eq.~\eqref{eq:sde}. No additional assumption or anticipation parameter is introduced.

States and controls are unbounded, coefficients are time independent, and we use risk-neutral expected costs. The equal, independent noise intensities isolate asymmetry originating in the optimized drift rather than an anisotropic diffusion matrix. We restrict attention to feedback Nash equilibria whose resulting dynamics are stable and admit a stationary probability distribution; the precise stability condition is specified after deriving the drift matrix. Since LQ differential games can have multiple feedback Nash equilibria \citelink{weeren1999}, we follow a selected stabilizing branch throughout (\applink{app:numerical}). Statements about finite $\tau$ refer to this branch, and short-foresight expansions are local asymptotic statements in $\tau$. Both states are taken even under time reversal; the entropy production reported here is a pathwise statistical irreversibility rate.

\subsection{Quadratic representation and Hamilton--Jacobi--Bellman equations}
Let $\bm e_1=(1,0)^T$, $\bm e_2=(0,1)^T$, $E_i=\bm e_i\bm e_i^T$, and write
\begin{equation}
 c_i(\bm x)=L_i(\bm x,0)
 =\tfrac12\bm x^TQ_i\bm x+\bm d_i^T\bm x+c_i^{(0)},
 \label{eq:qcost}
\end{equation}
where
\begin{align}
 Q_1&=\begin{pmatrix}a+b_1&-as_1\\-as_1&a\end{pmatrix},&
 Q_2&=\begin{pmatrix}a&-as_2\\-as_2&a+b_2\end{pmatrix},\label{eq:Q}\\
 \bm d_1&=(-b_1m_1,0)^T,&\bm d_2&=(0,-b_2m_2)^T,
 \qquad c_i^{(0)}=\tfrac12b_im_i^2.\label{eq:d}
\end{align}
With a quadratic value-function ansatz, including linear and constant terms,
\begin{equation}
 V_i(\bm x)=\tfrac12\bm x^TP_i\bm x+\bm h_i^T\bm x+k_i,
 \qquad P_i=P_i^T,
 \label{eq:value}
\end{equation}
we have $\nabla V_i=P_i\bm x+\bm h_i$ and $\nabla^2V_i=P_i$. The discounted stochastic Hamilton--Jacobi--Bellman (HJB) equations for mutually consistent policies are
\begin{equation}
 \rho V_i=\min_{u_i}\left\{c_i+\frac r2u_i^2
 +(\nabla V_i)^T(\bm e_i u_i+\bm e_j u_j^*)
 +D\tr P_i\right\}.
 \label{eq:hjb}
\end{equation}
The factor $D\tr P_i$ follows from It\^o's second-order term: the noise covariance is $2DI$, giving $(1/2)(2D)\tr\nabla^2 V_i$. Differentiation of the right-hand side with respect to $u_i$ yields $ru_i+\bm e_i^T(P_i\bm x+\bm h_i)=0$. Since $r>0$, the unique minimizing control, given the opponent's equilibrium rule, is
\begin{equation}
 u_i^*(\bm x)=-\frac1r\bm e_i^T(P_i\bm x+\bm h_i).
 \label{eq:policy}
\end{equation}

Insert Eq.~\eqref{eq:policy} for \emph{both} agents into Eq.~\eqref{eq:hjb}. The self-control terms combine into $-(P_i\bm x+\bm h_i)^TE_i(P_i\bm x+\bm h_i)/(2r)$, while the opponent's control contributes $-(P_i\bm x+\bm h_i)^TE_j(P_j\bm x+\bm h_j)/r$. Matching the quadratic and linear coefficients of $\bm x$ gives, respectively,
\begin{align}
 \rho P_i=Q_i-\frac1r\left(P_iE_iP_i+P_iE_jP_j+P_jE_jP_i\right),\label{eq:riccati}\\
 \rho\bm h_i=\bm d_i-\frac1r\left(P_iE_i\bm h_i+P_iE_j\bm h_j+P_jE_j\bm h_i\right).\label{eq:hlinear}
\end{align}
The constant terms determine $k_i$ through
\begin{equation}
 \rho k_i=c_i^{(0)}-\frac{\bm h_i^TE_i\bm h_i}{2r}
 -\frac{\bm h_i^TE_j\bm h_j}{r}+D\tr P_i.
 \label{eq:k}
\end{equation}
Equations~\eqref{eq:riccati} are coupled algebraic Riccati equations. They determine the feedback matrices, while the linear system~\eqref{eq:hlinear} determines the affine terms. Because $m_i$ appear only in $\bm d_i$, not $Q_i$, they cannot change the quadratic feedback coefficients on a fixed solution branch.

\subsection{Explicit six-equation system and effective OU dynamics}
Set
\begin{equation}
 P_1=\begin{pmatrix}A&C\\C&E\end{pmatrix},\qquad
 P_2=\begin{pmatrix}F&G\\G&H\end{pmatrix}.
 \label{eq:Pentries}
\end{equation}
Matrix multiplication in Eq.~\eqref{eq:riccati} produces six scalar equations:
\begin{align}
 \rho A &= a+b_1-\frac{A^2+2CG}{r},\nonumber\\
 \rho C &= -as_1-\frac{AC+CH+EG}{r},\nonumber\\
 \rho E &= a-\frac{C^2+2EH}{r},\nonumber\\
 \rho F &= a-\frac{G^2+2AF}{r},\nonumber\\
 \rho G &= -as_2-\frac{GH+FC+AG}{r},\nonumber\\
 \rho H &= a+b_2-\frac{H^2+2CG}{r}.
 \label{eq:six}
\end{align}
These six nonlinear equations are solved numerically. The coupling signs $s_i$ enter explicitly through the constant terms in the equations for $C$ and $G$. Writing $\bm h_1=(p_1,q_1)^T$ and $\bm h_2=(p_2,q_2)^T$, the equilibrium controls are
\begin{equation}
 u_1^*=-\frac{Ax_1+Cx_2+p_1}{r},\qquad
 u_2^*=-\frac{Gx_1+Hx_2+q_2}{r}.
 \label{eq:policies}
\end{equation}
Hence the closed-loop process is
\begin{align}
 d\bm x&=(-B\bm x+\bm f_0)\,dt+\sqrt{2D}\,d\bm W,\nonumber\\
 B&=\frac1r\begin{pmatrix}A&C\\G&H\end{pmatrix},&
 \bm f_0&=-\frac1r(p_1,q_2)^T.
 \label{eq:drift}
\end{align}
For a positive-stable matrix $B$ (all eigenvalues have positive real part), the stationary mean is $\bm\mu=B^{-1}\bm f_0$. With $\bm y=\bm x-\bm\mu$, the centered dynamics read
\begin{equation}
 d\bm y=-B\bm y\,dt+\sqrt{2D}\,d\bm W.
 \label{eq:centered}
\end{equation}
The centered dynamics therefore form a two-dimensional OU process. The preferred states $m_i$ affect the stationary mean through $\bm f_0$ but not $B$, the stationary covariance, or any of the irreversibility measures below (\applink{app:affine}). We define the effective drift asymmetry $\Delt=B_{12}-B_{21}=(C-G)/r$. Whether $\Delt$ vanishes is determined by the optimized response, not imposed in the interaction costs.

The parameter space can be reduced further. Rescaling $P_i=r\rho\,\tilde P_i$ turns Eq.~\eqref{eq:riccati} into $\tilde P_i=(\tau^2/r)Q_i-(\tilde P_iE_i\tilde P_i+\tilde P_iE_j\tilde P_j+\tilde P_jE_j\tilde P_i)$, where $Q_i/a$ depends only on $s_i$ and $\beta_i\equiv b_i/a$. All dimensionless ratios of entries of $B$, including the normalized quantities introduced in \reflink{sec:ou}, therefore depend only on $s_1$, $s_2$, $\beta_1$, $\beta_2$, and the dimensionless foresight parameter
\begin{equation}
 \theta\equiv\tau\sqrt{a/r}.
 \label{eq:theta}
\end{equation}
The timescale $\sqrt{r/a}$ balances control and interaction costs, so short foresight means $\tau\ll\sqrt{r/a}$. Equivalently, $\theta^2=a\tau^2/r$ compares the planning timescale $\tau$ with the closed-loop relaxation time $r/(a\tau)$ of nearly myopic agents (\applink{app:myopic}), and the short-foresight expansions of \reflink{sec:coord} and \reflink{sec:cc} are expansions in powers of $\theta^2$. All figures use $a=r=1$, for which $\theta=\tau$.

\section{Stationary current, irreversibility, and information flow}\label{sec:ou}
\subsection{Covariance and current}
Applying It\^o's product rule to $\bm y\bm y^T$ in Eq.~\eqref{eq:centered} gives the second-moment evolution
\begin{equation}
 \frac{d}{dt}\Sigma(t)=-B\Sigma(t)-\Sigma(t)B^T+2DI,
 \quad \Sigma(t)=\E[\bm y(t)\bm y(t)^T].
 \label{eq:sigmadyn}
\end{equation}
The stationary covariance is the unique positive-definite solution of the continuous Lyapunov equation
\begin{equation}
 B\Sigma+\Sigma B^T=2DI.
 \label{eq:lyapunov}
\end{equation}
It follows that the stationary density is $P_{\mathrm{st}}(\bm y)=(2\pi\sqrt{\det\Sigma})^{-1}\exp[-\bm y^T\Sigma^{-1}\bm y/2]$. The Fokker--Planck continuity equation, $\partial_tP=-\nabla\cdot\bm J$, gives
\begin{align}
 \bm J_{\mathrm{st}}&=-B\bm yP_{\mathrm{st}}-D\nabla P_{\mathrm{st}}
 =-(B-D\Sigma^{-1})\bm yP_{\mathrm{st}}.\label{eq:J}
\end{align}
Introduce the antisymmetric matrix
\begin{equation}
 \Omega=B\Sigma-DI=\frac12(B\Sigma-\Sigma B^T),
 \qquad \Omega^T=-\Omega.
 \label{eq:omega}
\end{equation}
Then $\bm J_{\mathrm{st}}=-\Omega\Sigma^{-1}\bm yP_{\mathrm{st}}$. In two dimensions write $B=\bigl(\begin{smallmatrix}B_{11}&B_{12}\\B_{21}&B_{22}\end{smallmatrix}\bigr)$ and $\Omega=\bigl(\begin{smallmatrix}0&w\\-w&0\end{smallmatrix}\bigr)$. From $B\Sigma=DI+\Omega$ and the symmetry of $\Sigma=B^{-1}(DI+\Omega)$, the off-diagonal entries obey $B_{22}w-B_{12}D=-B_{21}D-B_{11}w$. Thus
\begin{equation}
 w=D\frac{B_{12}-B_{21}}{B_{11}+B_{22}}=D\frac{\Delt}{\tr B}.
 \label{eq:w}
\end{equation}
The stationary current vanishes iff $w=0$, so for equal isotropic diffusion and time-even states, detailed balance holds precisely when $B_{12}=B_{21}$. The full covariance obtained from the same calculation is recorded in \applink{app:cov}.

\subsection{Exact stationary entropy production}
For even state variables and constant isotropic diffusion, the pathwise entropy-production rate (the stationary forward/reversed path relative-entropy rate) is \citelink{godreche2019,loos2020}
\begin{equation}
 \sigma=\int_{\mathbb R^2}\frac{\bm J_{\mathrm{st}}^T\bm J_{\mathrm{st}}}{D P_{\mathrm{st}}}\,d^2\bm y
 =\frac1D\tr(\Omega^T\Omega\Sigma^{-1}).
 \label{eq:epintegral}
\end{equation}
In two dimensions $\Omega^T\Omega=w^2I$. Taking the trace of Eq.~\eqref{eq:lyapunov} after multiplication on the right by $\Sigma^{-1}$ yields $\tr(\Sigma^{-1})=\tr B/D$. Substitution of this identity and Eq.~\eqref{eq:w} into Eq.~\eqref{eq:epintegral} gives the exact two-dimensional result
\begin{equation}
 \sigma=\frac{(B_{12}-B_{21})^2}{\tr B}
 =\frac{(C-G)^2}{r(A+H)}.
 \label{eq:ep}
\end{equation}
We also define the dimensionless \emph{normalized entropy production}
\begin{equation}
 \ep\equiv\frac{\sigma}{\tr B}
 =\left(\frac{B_{12}-B_{21}}{\tr B}\right)^2
 =\left(\frac{C-G}{A+H}\right)^2.
 \label{eq:epnormalized}
\end{equation}
Both quantities are independent of $D>0$ for this linear model: at fixed $B$, the stationary covariance is proportional to $D$, while the current field changes with the stationary probability density. The sign of $\Delt$ fixes the circulation orientation, whereas $\sigma$ and $\ep$ are nonnegative. We use $\ep$, the entropy produced per relaxation time $(\tr B)^{-1}$, to compare different planning timescales. This normalization matters because $B=O(\tau)$ as $\tau\to0$ (\applink{app:myopic}): nearly myopic agents barely move, so the unnormalized rate $\sigma$ partly reflects how fast the agents move rather than how irreversibly they do so. Consequently, $\sigma$ and $\ep$ have different exponents in the short-foresight limit; in the conformist--contrarian case of \reflink{sec:cc}, $\sigma$ even increases with $\tau$ at short foresight while $\ep$ decreases.

\subsection{Information flow between the agents}\label{sec:infoflow}
The sign of $\Delt$ fixes the orientation of the circulation but does not by itself state which agent's motion builds the correlation between the two states. Because the noises in Eq.~\eqref{eq:sde} are independent, the joint process is bipartite, and the rate of change of the mutual information between $y_1$ and $y_2$ splits into contributions from the dynamics of each coordinate \citelink{horowitz2014,loos2020}. The contribution of $y_1$ is
\begin{equation}
 \Iflow_1=\int_{\mathbb R^2}J_{\mathrm{st},1}\,
 \partial_{y_1}\ln\frac{P_{\mathrm{st}}(\bm y)}{P_{\mathrm{st}}^{(1)}(y_1)P_{\mathrm{st}}^{(2)}(y_2)}\,d^2\bm y,
 \label{eq:infodef}
\end{equation}
where $P_{\mathrm{st}}^{(1)}$ and $P_{\mathrm{st}}^{(2)}$ are the stationary marginal densities, and $\Iflow_2$ is defined analogously. In the stationary state the mutual information is constant, so $\Iflow_1+\Iflow_2=0$. A positive $\Iflow_1$ means that the motion of $y_1$ continuously builds correlation with $y_2$: agent 1 tracks agent 2. For the Gaussian stationary state (\applink{app:cov}),
\begin{equation}
 \Iflow_1=-\Iflow_2=w\,(\Sigma^{-1})_{12}=-\frac{w\,\Sigma_{12}}{\det\Sigma}.
 \label{eq:infoflow}
\end{equation}
Like $\sigma$, $\Iflow_1$ is independent of $D$; unlike $\sigma$, it is linear in $\Delt$ and carries its sign. We report it in the normalized form $\Iflow_1/\tr B$, the information built per relaxation time.

\section{Reciprocal coordination: emergent irreversibility}\label{sec:coord}
For $(s_1,s_2)=(+1,+1)$, the instantaneous costs have a common potential
\begin{equation}
 U(\bm x)=\frac a2(x_1-x_2)^2+
 \sum_{i=1}^2\frac{b_i}{2}(x_i-m_i)^2,
 \qquad \frac{\partial U}{\partial x_i}=\frac{\partial c_i}{\partial x_i}.
 \label{eq:potential}
\end{equation}
To leading order in $\tau$, the Nash feedback is the gradient flow $-(r\rho)^{-1}\nabla U$ (\applink{app:myopic}), which satisfies detailed balance. With foresight the Nash feedback need not remain a gradient drift. We derive its short-foresight asymmetry rather than assume it.

\subsection{Short-foresight expansion and scaling laws}\label{sec:expansion}
\begin{figure}[tbp]
\centering
\includegraphics[width=0.9\linewidth,height=0.48\textheight,keepaspectratio]{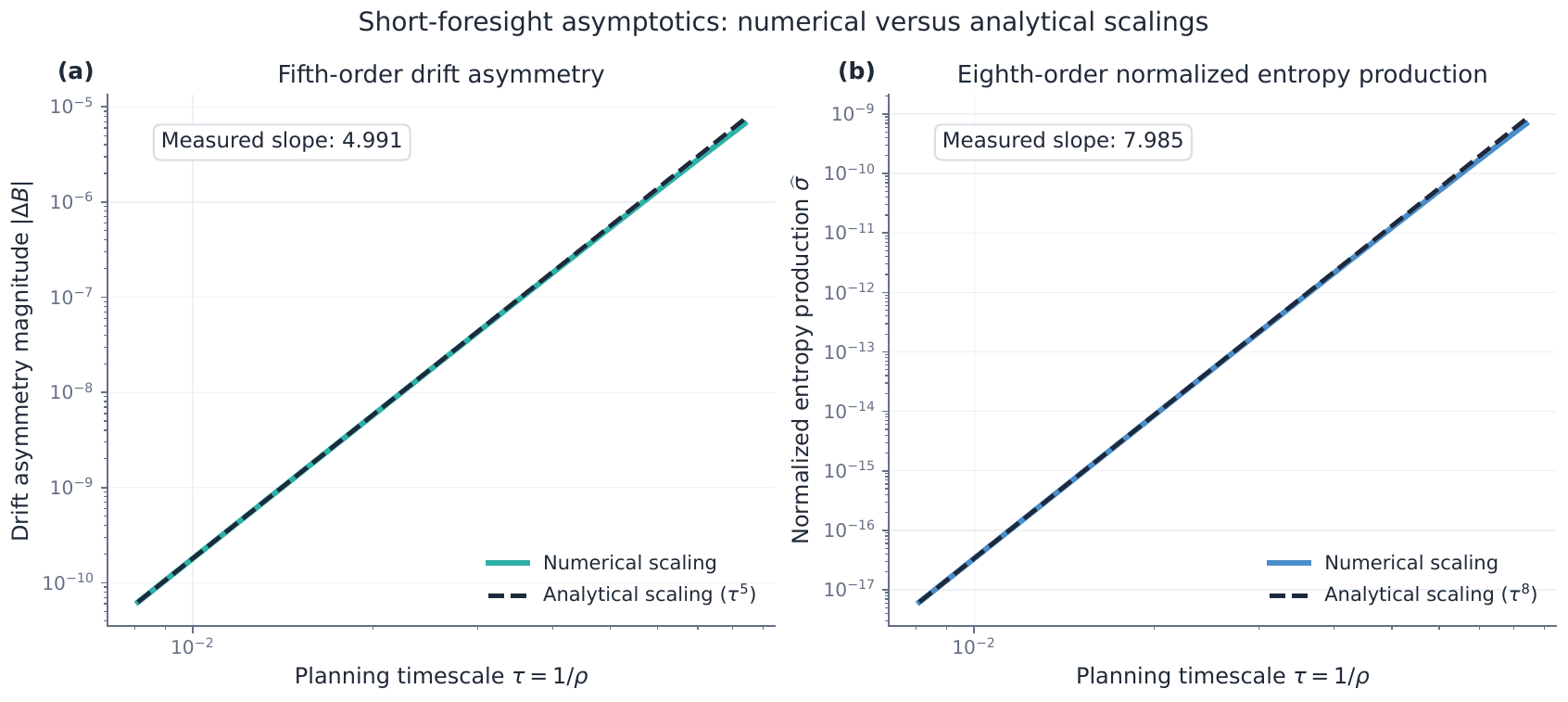}
\caption{\textbf{Fifth- and eighth-order short-foresight laws in reciprocal coordination.} Full-information Nash results (solid) compared with the leading analytical predictions (dashed) for $a=r=D=1$, $b_1=0.1$, $b_2=1$, and $(s_1,s_2)=(+1,+1)$. The fitted slopes are approximately $4.991$ for $|\Delt|$ and $7.985$ for $\ep$, against analytical exponents $5$ and $8$.}
\label{fig:coordscaling}
\end{figure}

Rewriting Eq.~\eqref{eq:riccati} as
\begin{equation}
 P_i=\tau Q_i-\frac{\tau}{r}
 (P_iE_iP_i+P_iE_jP_j+P_jE_jP_i),
 \label{eq:ricceps}
\end{equation}
shows that on the branch $P_i\to0$ as $\tau\to0$, the leading term is $P_i=\tau Q_i$. Iterating Eq.~\eqref{eq:ricceps} on the branch $P_i=O(\tau)$ gives an expansion in odd powers of $\tau$, with the first correction at cubic order:
\begin{equation}
 P_i=\tau Q_i
 -\frac{\tau^3}{r}
 \left(Q_iE_iQ_i+Q_iE_jQ_j+Q_jE_jQ_i\right)
 +O(\tau^5).
 \label{eq:Pexp}
\end{equation}
Evaluating the matrix entries using Eq.~\eqref{eq:six}, the coefficients through cubic order are
\begin{align}
 A&=(a+b_1)\tau-\frac{(a+b_1)^2+2a^2}{r}\tau^3+O(\tau^5),\nonumber\\
 C&=-a\tau+\frac{a(3a+b_1+b_2)}r\tau^3+O(\tau^5),\nonumber\\
 E&=a\tau-\frac{a(3a+2b_2)}r\tau^3+O(\tau^5),\nonumber\\
 F&=a\tau-\frac{a(3a+2b_1)}r\tau^3+O(\tau^5),\nonumber\\
 G&=-a\tau+\frac{a(3a+b_1+b_2)}r\tau^3+O(\tau^5),\nonumber\\
 H&=(a+b_2)\tau-\frac{(a+b_2)^2+2a^2}{r}\tau^3+O(\tau^5).
 \label{eq:cubiccoord}
\end{align}
In particular, $C-G=O(\tau^5)$: both the linear and cubic cross-coupling differences cancel. The fifth-order coefficient is obtained without calculating the complete fifth-order correction matrices. Subtract the $G$ equation from the $C$ equation in Eq.~\eqref{eq:six} (the constant terms are identical), and rearrange:
\begin{align}
 r\rho(C-G)&=-[AC+CH+EG-GH-FC-AG]\nonumber\\
 &=-C(A+H-F)-G(E-H-A).\label{eq:subtract}
\end{align}
Set $G=C-(C-G)$, collect the $(C-G)$ terms, and obtain the exact identity
\begin{equation}
 C-G=\frac{C(F-E)}{r\rho+A+H-E}.
 \label{eq:identity}
\end{equation}
From Eq.~\eqref{eq:cubiccoord},
\begin{align}
 F-E&=-\frac{2a(b_1-b_2)}r\tau^3+O(\tau^5),\nonumber\\
 C&=-a\tau+O(\tau^3),\qquad
 (r\rho+A+H-E)^{-1}=\frac{\tau}r+O(\tau^3).\label{eq:stepfive}
\end{align}
Combining these relations yields
\begin{align}
 C-G=\frac{2a^2(b_1-b_2)}{r^2}\tau^5+O(\tau^7),\label{eq:CminusG}\\
 \Delt=\frac{2a^2(b_1-b_2)}{r^3}\tau^5+O(\tau^7).\label{eq:delta5}
\end{align}
Thus discounted infinite-horizon Nash feedback preserves effective reciprocity through cubic order: the first nonzero term for $b_1\ne b_2$ is fifth order, proportional to the difference $b_1-b_2$. With $A+H=(2a+b_1+b_2)\tau+O(\tau^3)$, Eqs.~\eqref{eq:ep}--\eqref{eq:epnormalized} and \eqref{eq:CminusG} give
\begin{align}
 \ep=\frac{4a^4(b_1-b_2)^2}{r^4(2a+b_1+b_2)^2}\tau^8+O(\tau^{10}),\label{eq:hat8}\\
 \sigma=\frac{4a^4(b_1-b_2)^2}{r^5(2a+b_1+b_2)}\tau^9+O(\tau^{11}).\label{eq:sigma9}
\end{align}
In terms of Eq.~\eqref{eq:theta}, Eq.~\eqref{eq:hat8} reads $\ep=4\theta^8(\beta_1-\beta_2)^2/(2+\beta_1+\beta_2)^2+O(\theta^{10})$. These exponents hold for any fixed, unequal positive self-anchoring strengths, whose values affect the prefactors only. For $a=r=1$, $b_1=0.1$, and $b_2=1$, numerical fits over the short-foresight interval specified in \applink{app:numerical} give slopes approximately $4.991$ for $|\Delt|$ and $7.985$ for $\ep$ (\figref{fig:coordscaling}).

\subsection{Physical origin of the fifth-order law}\label{sec:mechanism}
The identity~\eqref{eq:identity} yields the fifth-order coefficient, but its structure also explains why the first two orders vanish and what sets the direction of the resulting current.

\emph{Cross-curvature criterion.} The identity~\eqref{eq:identity} does not rely on the specific form of Eq.~\eqref{eq:running}. Consider the same dynamics and control cost with arbitrary quadratic state costs $c_i$ that admit an instantaneous potential, that is, with a common cross term $(Q_1)_{12}=(Q_2)_{12}\equiv\kappa$, and let
\begin{equation}
 \gamma_1\equiv(Q_1)_{22},\qquad \gamma_2\equiv(Q_2)_{11}
 \label{eq:gammadef}
\end{equation}
denote each agent's cost curvature along the coordinate it does \emph{not} control. The constant terms of the $C$ and $G$ equations are then still identical, so Eq.~\eqref{eq:identity} holds unchanged, now with $C=\kappa\tau+O(\tau^3)$ and $F-E=(\gamma_2-\gamma_1)\tau+O(\tau^3)$. Hence
\begin{equation}
 \Delt=\frac{\kappa(\gamma_2-\gamma_1)}{r^2}\tau^3+O(\tau^5),
 \label{eq:crosscurv}
\end{equation}
so that $\ep\propto\tau^4$ whenever $\kappa(\gamma_2-\gamma_1)\ne0$.
Instantaneous incentives cannot see $\gamma_i$: the myopic control is proportional to $-\partial_{x_i}c_i$, which depends only on how $c_i$ varies with the agent's \emph{own} coordinate, and this is precisely why an instantaneous potential exists. With foresight, $\gamma_i$ matters, because an agent can steer its partner through the partner's feedback, and the value of doing so depends on how much the partner's state costs it. In Eq.~\eqref{eq:six}, this enters through the terms $EG$ and $FC$. For example, $FC$ combines agent 1's response to $x_2$ (gain $C$) with the curvature $F$ of agent 2's value function along $x_1$, which equals $\gamma_2\tau$ at leading order: agent 2 anticipates that when $x_2$ moves, agent 1 follows and reduces their mismatch. The more agent 2 cares about $x_1$, the more it relies on this accommodation and the less strongly it follows itself. Asymmetric reliance on the partner's accommodation is the only source of drift asymmetry in Eq.~\eqref{eq:identity}.

\begin{figure}[tbp]
\centering
\includegraphics[width=0.9\linewidth,height=0.48\textheight,keepaspectratio]{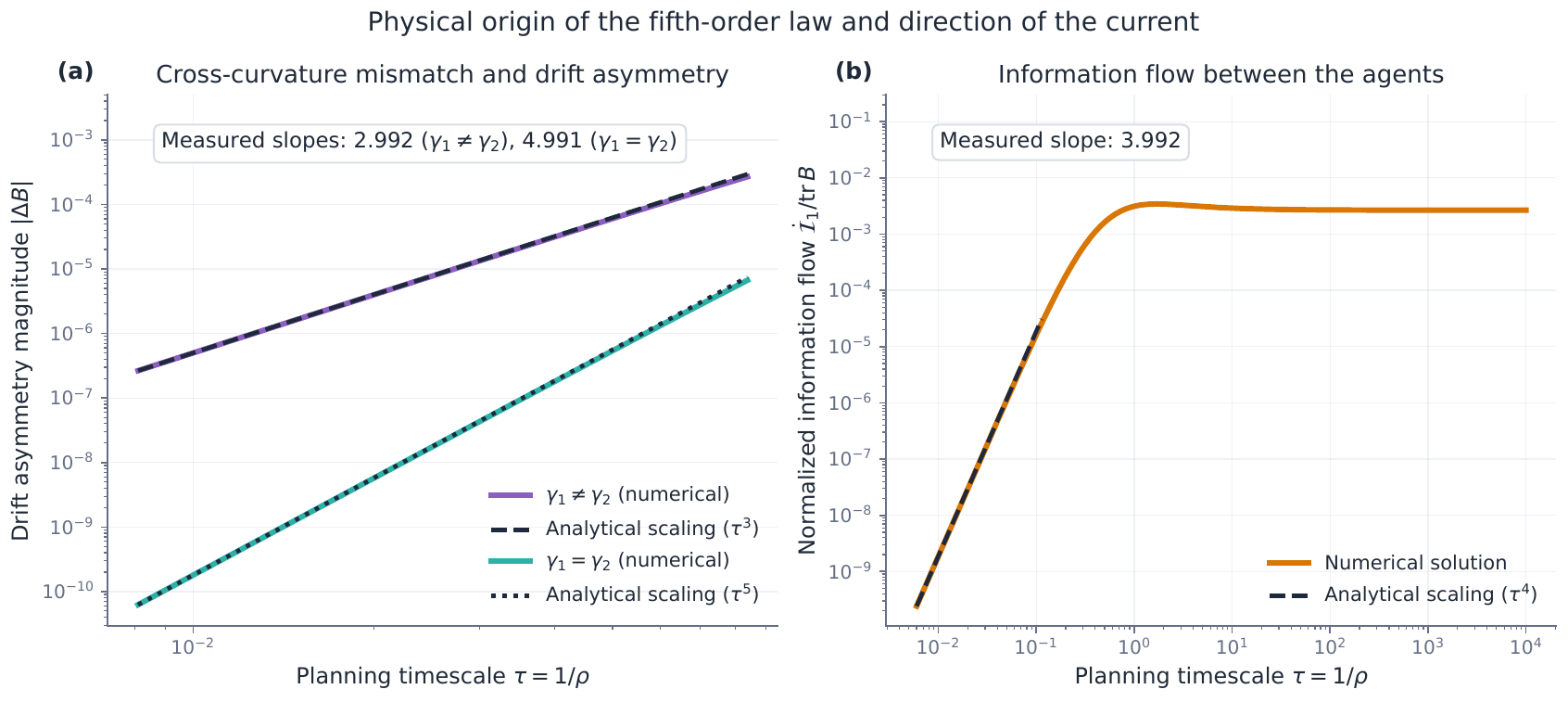}
\caption{\textbf{Physical origin of the fifth-order law and direction of the current.} (a) Drift-asymmetry magnitude $|\Delt|$ at short foresight for reciprocal coordination ($\gamma_1=\gamma_2=a$) and for the same game with the additional cost $g_1(x_2-n_1)^2/2$, $g_1=0.5$, in the objective of agent 1 ($\gamma_1=a+g_1\ne\gamma_2$), compared with the leading analytical laws, Eq.~\eqref{eq:delta5} ($\tau^5$, dotted) and Eq.~\eqref{eq:crosscurv} ($\tau^3$, dashed). (b) Normalized information flow $\Iflow_1/\tr B$ for reciprocal coordination over the full scanned range, compared with the leading analytical law, Eq.~\eqref{eq:info4} (dashed). Parameters: $a=r=D=1$, $b_1=0.1$, $b_2=1$, and $(s_1,s_2)=(+1,+1)$. The fitted slopes are approximately $2.992$ and $4.991$ in (a) and $3.992$ in (b), against analytical exponents $3$, $5$, and $4$.}
\label{fig:mechanism}
\end{figure}

\emph{Why the cubic term vanishes.} In Eq.~\eqref{eq:running}, $\gamma_1=\gamma_2=a$: the interaction cost $a(x_i-s_ix_j)^2/2$ gives each agent the same curvature along its partner's coordinate. The cubic cancellation in Eq.~\eqref{eq:cubiccoord} is therefore a structural property of the coordination cost. Adding a direct stake in the partner's state, $g_1(x_2-n_1)^2/2$ with $g_1>0$, to the cost of agent 1 gives $\gamma_1=a+g_1\ne\gamma_2$ and restores the cubic law, $\Delt=ag_1\tau^3/r^2+O(\tau^5)$, as confirmed numerically in \figref{fig:mechanism}(a). Because Eq.~\eqref{eq:crosscurv} does not involve $b_1$ and $b_2$, the cubic asymmetry persists even for equal self-anchoring strengths.

\emph{Why the asymmetry reappears at fifth order.} Foresight regenerates a cross-curvature mismatch dynamically. By Eq.~\eqref{eq:cubiccoord},
\begin{equation}
 E=a\tau-\frac{a^2}{r}\tau^3-\frac{2a(a+b_2)}{r}\tau^3+O(\tau^5),\qquad
 F=a\tau-\frac{a^2}{r}\tau^3-\frac{2a(a+b_1)}{r}\tau^3+O(\tau^5).
 \label{eq:EFsplit}
\end{equation}
The term $-a^2\tau^3/r$ is common to $E$ and $F$. At this order the anchoring strength $b_2$ enters $E$ only through the term $-2EH/r$ in Eq.~\eqref{eq:six}, which describes agent 2's own feedback relaxing $y_2$ at the rate $\nu_2\equiv H/r\simeq(a+b_2)\tau/r$. A displacement $\delta y_2$ costs agent 1 an amount $a\,\delta y_2^2/2$ per unit time only while it persists; since $\delta y_2^2$ decays as $e^{-2\nu_2t}$, its discounted weight is
\begin{equation}
 \int_0^\infty e^{-\rho t}e^{-2\nu_2t}\,dt
 =\tau\left[1-\frac{2(a+b_2)}{r}\tau^2+O(\tau^4)\right],
 \label{eq:persistence}
\end{equation}
which reproduces the last term of $E$ exactly. The same argument, with $\nu_1\equiv A/r\simeq(a+b_1)\tau/r$, gives the last term of $F$. A more strongly anchored partner's fluctuations are shorter lived and therefore matter less. Unequal anchoring thus produces unequal \emph{effective} cross-curvatures, $F-E$ in Eq.~\eqref{eq:stepfive}. Inserting the effective mismatch $(\gamma_2-\gamma_1)_{\mathrm{eff}}=(F-E)/\tau=-2a(b_1-b_2)\tau^2/r$ into Eq.~\eqref{eq:crosscurv} with $\kappa=-a$ reproduces Eq.~\eqref{eq:delta5}. The fifth power is therefore the cubic mechanism acting on a mismatch that itself first appears at relative order $\tau^2$. In terms of Eq.~\eqref{eq:theta},
\begin{equation}
 \frac{\Delt}{\tr B}
 =\frac{2\theta^4(\beta_1-\beta_2)}{2+\beta_1+\beta_2}+O(\theta^6),
 \label{eq:ratio4}
\end{equation}
which is second order in $\theta^2$: one order generates the persistence mismatch and one converts it into unequal feedback gains.

\emph{Direction of the current.} For $b_1<b_2$, Eq.~\eqref{eq:EFsplit} gives $F>E$: the strongly anchored agent 2 relies more on agent 1's accommodation, because agent 1's fluctuations persist longer, and therefore follows less. Hence $|B_{12}|>|B_{21}|$ and $\Delt<0$: the weakly anchored agent responds more strongly to its partner. The information flow of \reflink{sec:infoflow} quantifies this asymmetry. Since $\Sigma=DB^{-1}$ when $B$ is symmetric, $\Sigma^{-1}=B/D+O(\Delt)$, and Eq.~\eqref{eq:infoflow} gives $\Iflow_1=B_{12}\Delt/\tr B+O(\Delt^2)$, so that
\begin{equation}
 \frac{\Iflow_1}{\tr B}
 =\frac{2\theta^4(\beta_2-\beta_1)}{(2+\beta_1+\beta_2)^2}+O(\theta^6).
 \label{eq:info4}
\end{equation}
The weakly anchored agent is thus the tracking side, and the normalized information flow appears at order $\theta^4$, four orders before $\ep\propto\theta^8$. \figref{fig:mechanism}(b) confirms Eq.~\eqref{eq:info4} and shows that $\Iflow_1>0$ over the full scanned range for $b_1=0.1$ and $b_2=1$; exchanging $b_1$ and $b_2$ relabels the agents and reverses the sign.

\subsection{Finite-foresight behavior and self-anchoring heterogeneity}\label{sec:finite}
\begin{figure}[tbp]
\centering
\includegraphics[width=0.9\linewidth,height=0.48\textheight,keepaspectratio]{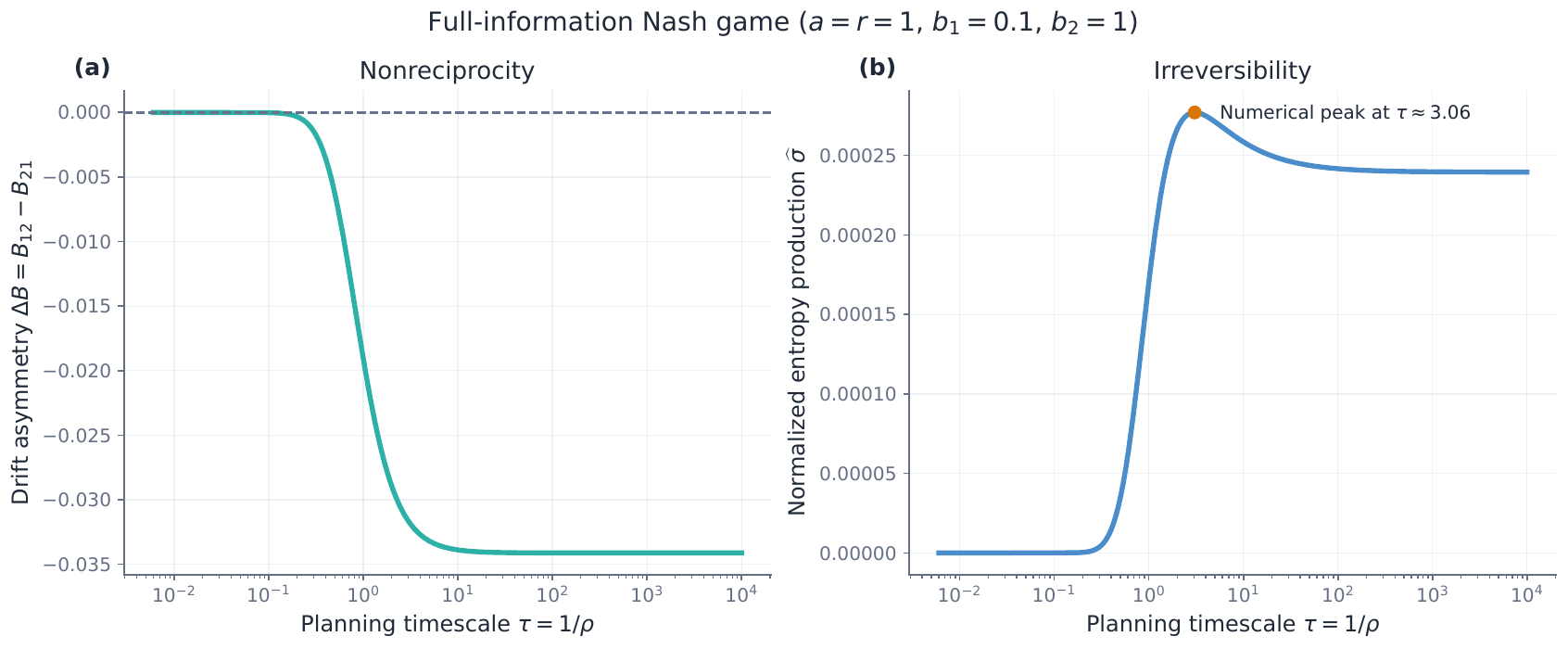}
\caption{\textbf{Irreversibility over the full scanned planning-timescale range in reciprocal coordination.} (a) Drift asymmetry $\Delt=B_{12}-B_{21}$ and (b) normalized entropy production $\ep$ versus $\tau$, with $a=r=D=1$, $b_1=0.1$, $b_2=1$. The dashed zero line in (a) represents the short-foresight limit where $\Delt\to0$; at large $\tau$, the asymmetry approaches a negative numerical plateau ($\Delt\simeq-0.034$). The marked peak and plateau are numerical properties of the selected Nash branch, not assumed analytical limits.}
\label{fig:coordfull}
\end{figure}

Beyond the short-foresight regime, $\ep$ rises from its eighth-order onset, reaches a numerical maximum $\ep\simeq2.77\times10^{-4}$ near $\tau\simeq3.06$, and then decreases toward a plateau over the scanned range (\figref{fig:coordfull}). The corresponding $\Delt$ is negative for this parameter ordering, consistent with the direction identified in \reflink{sec:mechanism}.

\begin{figure}[H]
\centering
\includegraphics[width=1.0\linewidth,height=0.47\textheight,keepaspectratio]{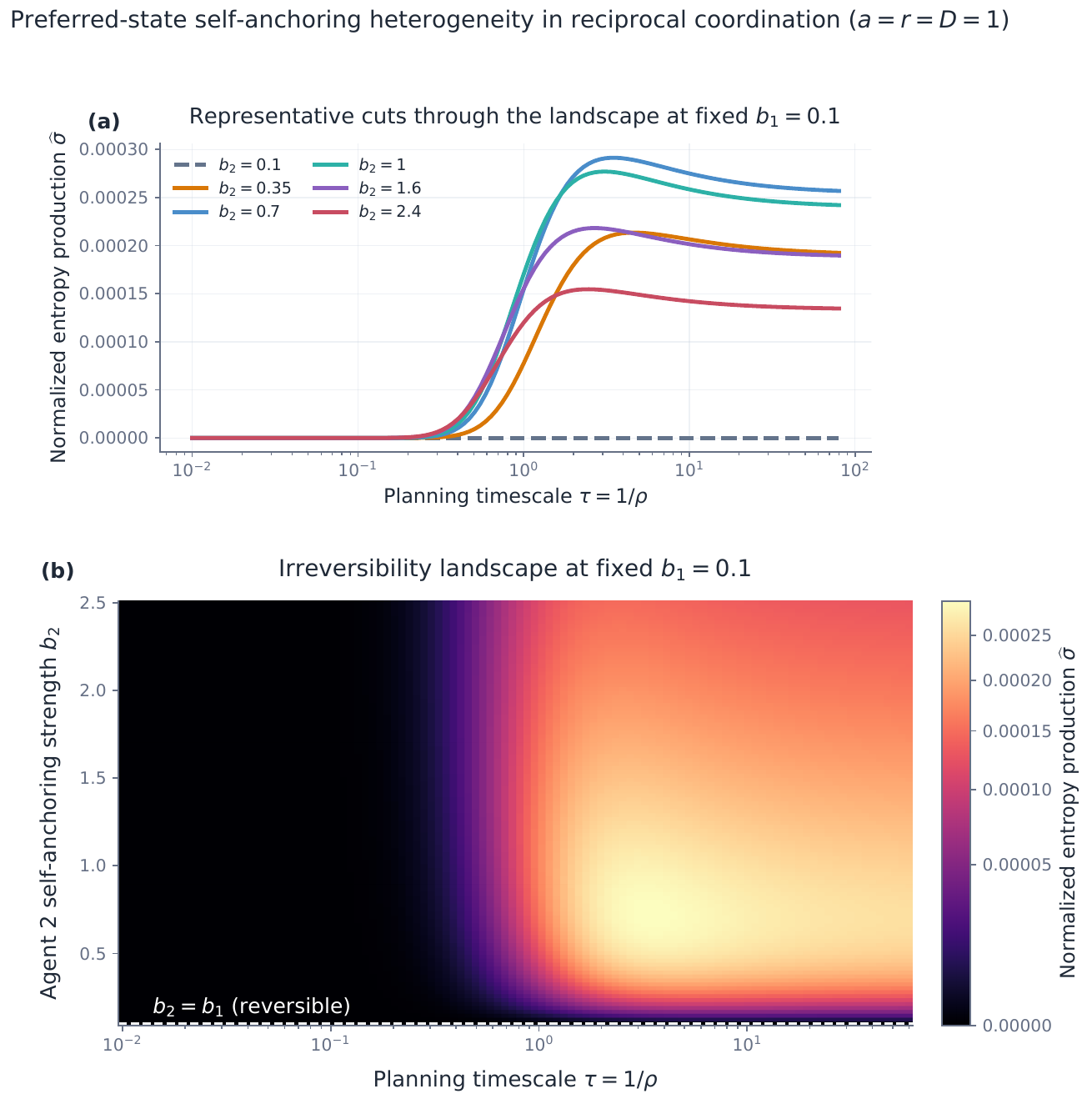}
\caption{\textbf{Preferred-state self-anchoring heterogeneity in reciprocal coordination.} (a) Normalized entropy production $\ep$ versus $\tau$ at $a=r=D=1$ and fixed $b_1=0.1$, for selected values of the agent-2 self-anchoring strength $b_2$. The $b_2=0.1$ branch is reversible. (b) Numerical landscape of $\ep$ as a function of $\tau$ and $b_2$ at the same fixed $b_1$. The dashed line marks $b_2=b_1$. Panel (a) resolves the individual curves and their nonmonotonic dependence on $b_2$; panel (b) displays the two-parameter landscape.}
\label{fig:coordhetero}
\end{figure}

At fixed $b_1=0.1$, increasing $b_2$ changes both the asymmetry and the overall relaxation rate; it therefore need not increase $\ep$ monotonically. For example, the peak value is larger at $b_2=0.7$ than at $b_2=2.4$, despite the greater difference $|b_2-b_1|$ in the latter case. We display one-dimensional scans together with a two-parameter landscape in \figref{fig:coordhetero}. The exactly symmetric line $b_1=b_2$ is reversible on the exchange-symmetric Nash branch, as shown in \applink{app:symmetry}.

\FloatBarrier

\section{Conformist--contrarian coupling: suppression of normalized entropy production}\label{sec:cc}
Now take $(s_1,s_2)=(+1,-1)$. Relative to reciprocal coordination, only the off-diagonal entries of $Q_2$ change sign: $(Q_2)_{12}=(Q_2)_{21}=+a$. Thus agent 1 favors $x_1=x_2$, whereas agent 2 favors $x_2=-x_1$ in its interaction cost, before balancing self-anchoring and control costs. Since $(Q_1)_{12}\ne(Q_2)_{12}$, the instantaneous costs admit no potential, and the short-foresight drift is already nonsymmetric: $C=-a\tau+O(\tau^3)$ and $G=+a\tau+O(\tau^3)$.

\subsection{Short-foresight expansion}
Substituting these signs into Eq.~\eqref{eq:six}, and proceeding as in \reflink{sec:expansion}, gives
\begin{align}
 A&=(a+b_1)\tau+\frac{2a^2-(a+b_1)^2}{r}\tau^3+O(\tau^5),\nonumber\\
 H&=(a+b_2)\tau+\frac{2a^2-(a+b_2)^2}{r}\tau^3+O(\tau^5),\nonumber\\
 C&=-a\tau+\frac{a(a+b_1+b_2)}r\tau^3+O(\tau^5),\nonumber\\
 G&=+a\tau-\frac{a(a+b_1+b_2)}r\tau^3+O(\tau^5).
 \label{eq:ccseries}
\end{align}
Therefore,
\begin{align}
 C-G&=-2a\tau+\frac{2a(a+b_1+b_2)}r\tau^3+O(\tau^5),\nonumber\\
 A+H&=(2a+b_1+b_2)\tau
 +\frac{4a^2-(a+b_1)^2-(a+b_2)^2}{r}\tau^3+O(\tau^5).
 \label{eq:ccratio}
\end{align}
Let $\Lambda=2a+b_1+b_2$, $M=a+b_1+b_2$, and $N=4a^2-(a+b_1)^2-(a+b_2)^2$. Inserting Eq.~\eqref{eq:ccratio} into $\ep=[(C-G)/(A+H)]^2$ and expanding $(1+z)^{-2}=1-2z+O(z^2)$ yields
\begin{equation}
 \ep=\frac{4a^2}{\Lambda^2}
 -\frac{8a^2(M\Lambda+N)}{r\Lambda^3}\tau^2+O(\tau^4).
 \label{eq:ccKinter}
\end{equation}
Since $M\Lambda+N=4a^2+a(b_1+b_2)+2b_1b_2>0$, we obtain
\begin{equation}
 \ep=\frac{4a^2}{(2a+b_1+b_2)^2}
 -\frac{8a^2[4a^2+a(b_1+b_2)+2b_1b_2]}{r(2a+b_1+b_2)^3}\tau^2
 +O(\tau^4).
 \label{eq:ccshort}
\end{equation}
In terms of Eq.~\eqref{eq:theta}, $\ep=4/(2+\beta_1+\beta_2)^2-8\theta^2(4+\beta_1+\beta_2+2\beta_1\beta_2)/(2+\beta_1+\beta_2)^3+O(\theta^4)$. Thus the conformist--contrarian case has nonzero \emph{normalized} entropy production as $\tau\to0^+$, and increasing $\tau$ initially decreases it. The reduction from the short-foresight limit is quadratic, as illustrated numerically in \figref{fig:ccscans}(c) for $a=r=D=1$ and $b_1=b_2=0.5$.

\subsection{Long-foresight limit for equal self-anchoring strengths}\label{sec:cclong}
\begin{figure}[H]
\centering
\includegraphics[width=0.99\linewidth,height=0.57\textheight,keepaspectratio]{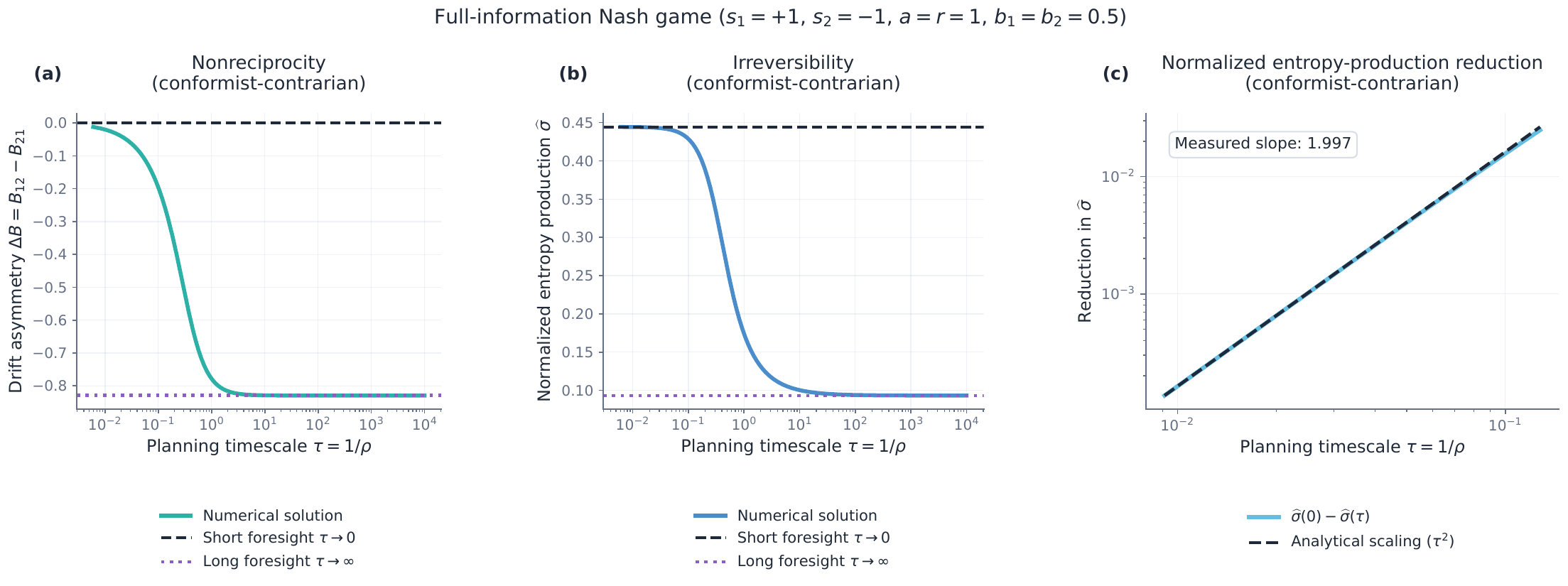}
\caption{\textbf{Foresight-dependent nonreciprocity and irreversibility in conformist--contrarian coupling.} For $(s_1,s_2)=(+1,-1)$, $a=r=D=1$, and $b_1=b_2=0.5$: (a) effective drift asymmetry $\Delt=B_{12}-B_{21}$ versus $\tau$, with its short- and long-foresight limits; (b) normalized entropy production $\ep$, with its exact short-foresight limit $4/9$ and long-foresight limit $z_*^2\simeq0.093286$; (c) reduction in normalized entropy production, $\ep(0)-\ep(\tau)$, where $\ep(0)$ denotes the $\tau\to0^+$ limit, compared with the analytical quadratic scaling from Eq.~\eqref{eq:ccshort} (fitted exponent approximately $1.997$).}
\label{fig:ccscans}
\end{figure}

For $b_1=b_2=b$, the conformist--contrarian game admits a symmetry-compatible branch with
\begin{equation}
 P_2=S_{\mathrm{cc}}P_1S_{\mathrm{cc}}^T,
 \qquad S_{\mathrm{cc}}=\begin{pmatrix}0&-1\\1&0\end{pmatrix},
 \qquad F=E,\quad G=-C,\quad H=A.
 \label{eq:ccsym}
\end{equation}
The six equations reduce to
\begin{align}
 \rho A&=a+b-\frac{A^2-2C^2}{r},\nonumber\\
 \rho C&=-a-\frac{C(2A-E)}r,\nonumber\\
 \rho E&=a-\frac{C^2+2AE}{r}.
 \label{eq:cc3}
\end{align}
On a finite stabilizing branch as $\rho\to0^+$, these imply
\begin{equation}
 A^2-2C^2=r(a+b),\quad C(2A-E)=-ra,
 \quad C^2+2AE=ra.
 \label{eq:cczero}
\end{equation}
Set $z=C/A$. The drift $B=r^{-1}\bigl(\begin{smallmatrix}A&C\\-C&A\end{smallmatrix}\bigr)$ has eigenvalues $(A\pm iC)/r$ and is therefore stable for $A>0$. The first relation gives $A^2=r(a+b)/(1-2z^2)$; the second gives $E=2A+ra/C$. Insert both into the third relation to obtain
\begin{equation}
 (a+b)z(z^2+4)=a(z-2)(1-2z^2),
 \label{eq:ccpolyinter}
\end{equation}
so $z$ obeys the cubic
\begin{equation}
 f(z)\equiv(3a+b)z^3-4az^2+(3a+4b)z+2a=0.
 \label{eq:cccubic}
\end{equation}
This cubic has exactly one real root $z_*$: the derivative $f'$ has discriminant $-4(11a^2+45ab+12b^2)<0$, so $f$ is strictly increasing. Moreover, $f(0)=2a>0$, $f(-1/\sqrt2)=-9(a+b)/(2\sqrt2)<0$, and
\begin{equation}
 f\!\left(-\frac{a}{a+b}\right)=-\frac{a(8a^3+9a^2b+5ab^2+2b^3)}{(a+b)^3}<0,
 \label{eq:ccfvalue}
\end{equation}
so that $\max\{-1/\sqrt2,\,-a/(a+b)\}<z_*<0$. The first bound ensures $A^2=r(a+b)/(1-2z_*^2)>0$. Since $\ep=[(C-G)/(A+H)]^2=(C/A)^2$, the limit is
\begin{equation}
 \lim_{\tau\to\infty}\ep=z_*^2<\frac{a^2}{(a+b)^2}=\lim_{\tau\to0^+}\ep,
 \label{eq:cclimit}
\end{equation}
where the inequality follows from the second bound and the right-hand side is Eq.~\eqref{eq:ccshort} at $\tau\to0^+$ with $b_1=b_2=b$. Hence, whenever the symmetry-compatible branch has a finite stabilizing long-foresight limit, this limit is unique and lies strictly below the short-foresight value for all $a,b>0$. For $a=r=1$ and $b=0.5$, numerical solution of Eq.~\eqref{eq:cccubic} gives $z_*\simeq-0.305427$ and $z_*^2\simeq0.093286$, compared with $\ep\to4/9$ as $\tau\to0^+$. The numerical crossover is displayed in \figref{fig:ccscans}.

On this branch, $B+B^T=(2A/r)I$ for every $\tau$, so Eq.~\eqref{eq:lyapunov} is solved by $\Sigma=(rD/A)I$. Then $\Sigma_{12}=0$, and Eq.~\eqref{eq:infoflow} gives $\Iflow_1=0$: the stationary mutual-information flow vanishes, and the circulation is a pure rotation that builds no equal-time correlation between the agents, although $\ep$ remains of order unity. Time-lagged measures of information transfer need not vanish; $\Iflow_1$ refers only to the equal-time mutual information of Eq.~\eqref{eq:infodef}. Entropy production and information flow thus distinguish the two coupling classes; in heterogeneous reciprocal coordination, by contrast, the normalized information flow appears four orders in $\tau$ before the normalized entropy production.

\FloatBarrier

\section{Discussion and conclusions}\label{sec:discussion}

Our results show that forward-looking optimization can change the reversibility of two-agent stochastic dynamics even when the agents' immediate interaction preferences remain unchanged. In reciprocal coordination, the instantaneous costs admit the common potential $U$ of Eq.~\eqref{eq:potential}, and the myopic Nash feedback is its gradient flow. Foresight breaks this structure through a single channel: each agent's response becomes sensitive to how much its partner's state costs it, which instantaneous incentives cannot see. The coordination cost equalizes these cross-curvatures, so the cubic contribution to the drift asymmetry vanishes. Unequal self-anchoring makes the partners' fluctuations persist for different times and regenerates an effective cross-curvature mismatch at relative order $\tau^2$. This produces the fifth-order asymmetry of Eq.~\eqref{eq:delta5}, the $\tau^8$ law for $\ep$, and the $\tau^9$ law for $\sigma$. The same mechanism fixes the direction of the current: the weakly anchored agent follows its partner more strongly and is the side that builds correlation, with a normalized information flow of order $\tau^4$. Beyond short foresight, $\ep$ depends nonmonotonically on both $\tau$ and the anchoring heterogeneity (\figrefs{fig:coordfull}{fig:coordhetero}).

In conformist--contrarian coupling, the instantaneous costs admit no potential and the Nash drift is nonreciprocal already at leading order. Increasing foresight initially reduces $\ep$ by an amount proportional to $\tau^2$, Eq.~\eqref{eq:ccshort}, and for equal self-anchoring strengths the long-foresight limit lies strictly below the short-foresight value, Eq.~\eqref{eq:cclimit}. Here the magnitude of the drift asymmetry grows with foresight, but its magnitude relative to the overall relaxation rate decreases, so the dynamics become less irreversible per relaxation time. For equal anchoring, the circulation is a pure rotation for which the stationary mutual-information flow vanishes. This interpretation is an intuitive reading of the calculated feedback dynamics, and not a general prediction about how real agents resolve disagreement.

\begin{figure}[H]
\centering
\includegraphics[width=0.96\linewidth,height=0.48\textheight,keepaspectratio]{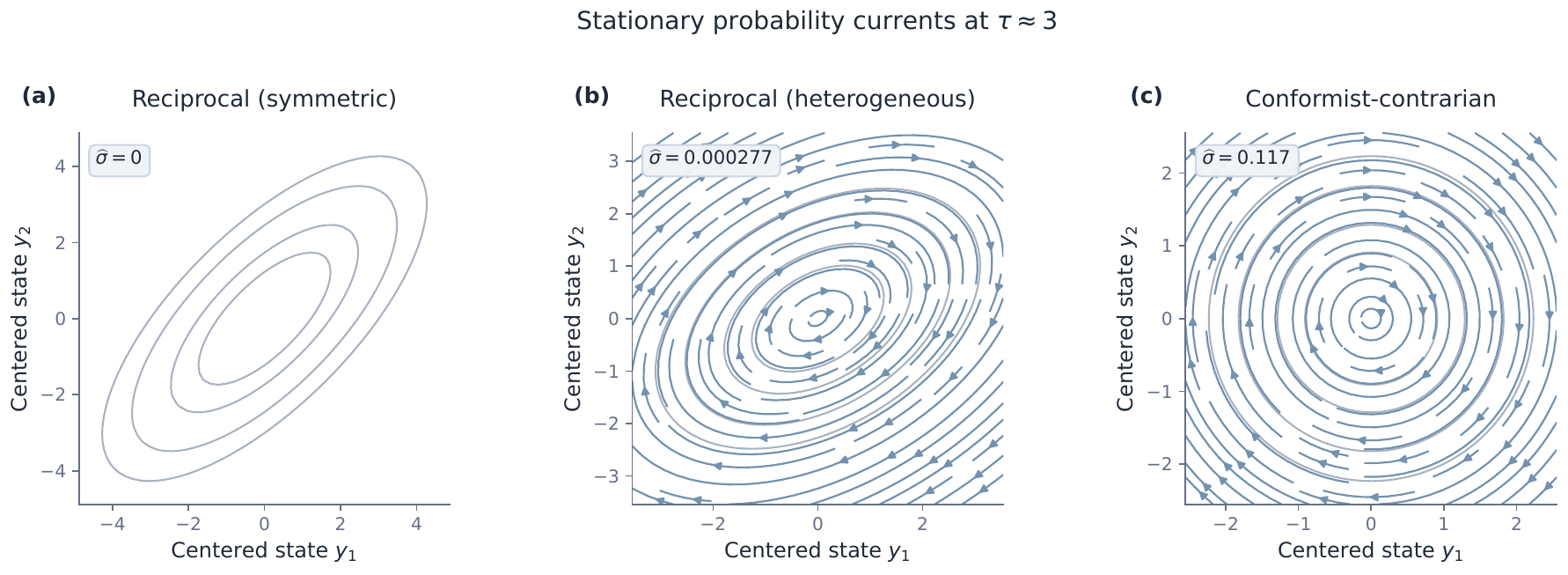}
\caption{\textbf{Stationary probability currents and detailed balance.} Level contours of the centered stationary Gaussian distributions at $\tau\simeq3$ and $a=r=D=1$. (a) Exchange-symmetric reciprocal coordination ($s_1=s_2=+1$, $b_1=b_2=0.1$): detailed balance holds and $\bm J_{\mathrm{st}}=0$. (b) Heterogeneous reciprocal coordination ($s_1=s_2=+1$, $b_1=0.1$, $b_2=1$): nonzero circulating current. (c) Conformist--contrarian coupling ($s_1=+1$, $s_2=-1$, $b_1=b_2=0.5$): nonzero circulating current. Streamlines in panels (b) and (c) represent $\bm J_{\mathrm{st}}/P_{\mathrm{st}}$; their speeds are separately normalized for visibility and cannot be directly compared by arrow length or density.}
\label{fig:currents}
\end{figure}

The corresponding stationary probability currents are illustrated in \figref{fig:currents}. Panel (a) provides a reversible reference, and panels (b) and (c) show the currents generated by heterogeneous reciprocal coordination and conformist--contrarian coupling. Because the centered dynamics are OU processes, all three stationary distributions are Gaussian, and elliptical density contours do not by themselves indicate irreversibility. The streamlines therefore show the local probability-current velocity $\bm J_{\mathrm{st}}/P_{\mathrm{st}}$, rather than the deterministic drift or individual stochastic trajectories; this removes the Gaussian density dependence of the current magnitude without changing the streamline geometry. The physical magnitudes of the two currents must be compared through $\ep$, Eq.~\eqref{eq:epnormalized}.

The mechanism is not specific to the present cost functions. Equation~\eqref{eq:crosscurv} applies to any quadratic state costs with an instantaneous potential, for the dynamics and control cost considered here, and identifies the cross-curvature mismatch as the quantity that foresight converts into nonreciprocity. More broadly, it indicates that for feedback controllers that anticipate each other, reciprocity of the instantaneous objectives does not guarantee detailed balance; what matters is how strongly each controller's objective depends on the coordinates it does not actuate. The model can be realized directly with two colloidal particles in feedback traps \citelink{jun2012}. Each trap applies a virtual force $\zeta u_i(x_1,x_2)$ computed from the measured positions by its own discounted LQ controller, so that Eq.~\eqref{eq:sde} holds with $D=k_BT/\zeta$, where $\zeta$ is the friction coefficient. The drift matrix, and hence $\Delt$, $\sigma$, and $\Iflow_1$, can be obtained from the stationary two-time correlations of the measured trajectories. In such a setting, $\sigma$ counts only the irreversibility of the controlled coordinates; the measurement and feedback channel carries its own entropy budget \citelink{cheung2026}. Feedback delay and measurement noise, neglected here, would add effective memory to the drift.

Several limitations should be kept in mind. The model contains only two continuous, unbounded state variables, with quadratic costs, equal isotropic noise, and full-state observation. The numerical calculations verify the Riccati equations, closed-loop stability, and independent best responses at sampled parameter values on the selected branch, but they do not establish global uniqueness of the Nash equilibrium or global monotonicity in $\tau$. The independence of $\sigma$, $\ep$, and $\Iflow_1$ from $D$ relies on the linear dynamics and isotropic diffusion. Unequal noise intensities would break detailed balance even for a symmetric drift, generically producing normalized irreversibility already as $\tau\to0^+$ that could mask the $\tau^8$ effect identified here. The behavior may also differ with correlated noise, nonlinear costs, or partial observation. Investigating these extensions, together with possible changes between equilibrium branches, would help determine how broadly the mechanisms identified here persist.

A final comparison isolates the role of individual optimization. A cooperative planner minimizing $J_1+J_2$ with the same dynamics, costs, and noise yields a symmetric drift matrix and hence vanishing stationary current and entropy production, even where the Nash dynamics are irreversible (\applink{app:planner}). The difference from the Nash results is therefore not simply whether the agents look ahead, but whether their future costs are optimized separately or jointly.

Taken together, these results demonstrate that foresight does not have a single, predetermined effect on stationary irreversibility. In reciprocal coordination, individually optimized feedback generates nonreciprocal dynamics from reciprocal preferences, through a cross-curvature mismatch that is invisible to instantaneous incentives; in conformist--contrarian coupling, it reduces normalized irreversibility despite an already asymmetric interaction. Because its stationary observables follow exactly from the Nash feedback, the model provides a quantitative connection between strategic optimization and stochastic thermodynamics: it identifies the quantity through which anticipation generates nonreciprocity, explains the resulting fifth-order law physically, and predicts the direction of the current through the information flow.

\section*{Code availability}

The numerical code, figure-generation script, and figures associated with this work are publicly available on GitHub at \url{https://github.com/arpand2004/foresight-entropy-production-game} and permanently archived on Zenodo at \doilink{10.5281/zenodo.24351245}. See \applink{app:numerical} for more details.

\section*{Declaration of AI use}

The research question, model, and approach are the author's own. Claude (Anthropic; Opus 5.5) was used to develop and debug the numerical code, to assist with parts of the analytical calculations, and to assist with manuscript preparation. ChatGPT (OpenAI; GPT-6) was used to cross-check calculations and to assist with formatting and presentation. The author reviewed and verified all analytical results, numerical code, and text, and takes full responsibility for the results and conclusions.

\appendix
\section*{Appendices}
\section{Explicit covariance, stationary-current, and information-flow identities}\label{app:cov}
For $B=\bigl(\begin{smallmatrix}B_{11}&B_{12}\\B_{21}&B_{22}\end{smallmatrix}\bigr)$ and $\Sigma=\bigl(\begin{smallmatrix}\Sigma_{11}&\Sigma_{12}\\\Sigma_{12}&\Sigma_{22}\end{smallmatrix}\bigr)$, Eq.~\eqref{eq:lyapunov} gives
\begin{align}
 B_{11}\Sigma_{11}+B_{12}\Sigma_{12}&=D,\nonumber\\
 B_{21}\Sigma_{11}+(B_{11}+B_{22})\Sigma_{12}+B_{12}\Sigma_{22}&=0,\nonumber\\
 B_{21}\Sigma_{12}+B_{22}\Sigma_{22}&=D.
 \label{eq:covscalar}
\end{align}
Equivalently $B\Sigma=DI+\Omega$, where $\Omega=\bigl(\begin{smallmatrix}0&w\\-w&0\end{smallmatrix}\bigr)$. Since $B^{-1}=(\det B)^{-1}\bigl(\begin{smallmatrix}B_{22}&-B_{12}\\-B_{21}&B_{11}\end{smallmatrix}\bigr)$, direct multiplication gives
\begin{equation}
 \Sigma=\frac1{\det B}\begin{pmatrix}
 DB_{22}+B_{12}w&B_{22}w-DB_{12}\\
 -DB_{21}-B_{11}w&DB_{11}-B_{21}w
 \end{pmatrix}.
 \label{eq:covtemp}
\end{equation}
Symmetry of $\Sigma$ implies $(\tr B)w=D(B_{12}-B_{21})$, whence
\begin{equation}
 \Sigma=\frac{D}{(\tr B)(\det B)}\begin{pmatrix}
 B_{22}\tr B+B_{12}\Delt&-(B_{11}B_{12}+B_{22}B_{21})\\
 -(B_{11}B_{12}+B_{22}B_{21})&B_{11}\tr B-B_{21}\Delt
 \end{pmatrix}.
 \label{eq:covexplicit}
\end{equation}
As an alternative check on Eq.~\eqref{eq:ep}, use $\bm J_{\mathrm{st}}=-\Omega\Sigma^{-1}\bm yP_{\mathrm{st}}$ in Eq.~\eqref{eq:epintegral}; because $\Omega^T\Omega=w^2I$ and $\tr(\Sigma^{-1})=\tr B/D$, one obtains $\sigma=w^2\tr B/D^2=\Delt^2/\tr B$.

For the information flow, the first component of the stationary current is $J_{\mathrm{st},1}=-w(\Sigma^{-1}\bm y)_2P_{\mathrm{st}}$. The stationary marginal $P_{\mathrm{st}}^{(1)}$ is Gaussian with variance $\Sigma_{11}$, and $P_{\mathrm{st}}^{(2)}$ does not depend on $y_1$, so $\partial_{y_1}\ln[P_{\mathrm{st}}/(P_{\mathrm{st}}^{(1)}P_{\mathrm{st}}^{(2)})]=-(\Sigma^{-1}\bm y)_1+y_1/\Sigma_{11}$. Using $\E[\Sigma^{-1}\bm y\bm y^T\Sigma^{-1}]=\Sigma^{-1}$ and $\E[\Sigma^{-1}\bm y\bm y^T]=I$ in Eq.~\eqref{eq:infodef} gives
\begin{equation}
 \Iflow_1=w(\Sigma^{-1})_{12}-\frac{w}{\Sigma_{11}}\,(I)_{21}=w(\Sigma^{-1})_{12},
 \label{eq:infoderiv}
\end{equation}
which is Eq.~\eqref{eq:infoflow}. The same calculation with $J_{\mathrm{st},2}=w(\Sigma^{-1}\bm y)_1P_{\mathrm{st}}$ gives $\Iflow_2=-w(\Sigma^{-1})_{12}$. The off-diagonal covariance $\Sigma_{12}$ follows from Eq.~\eqref{eq:covexplicit}.

\section{Analytical consistency checks}\label{app:checks}
\subsection{No coupling: \texorpdfstring{$a=0$}{a=0}}\label{app:uncoupled}
When $a=0$, the state costs decouple: $c_i=b_i(x_i-m_i)^2/2$. A solution of Eq.~\eqref{eq:six} has $C=G=E=F=0$ and
\begin{equation}
 A^2+r\rho A-rb_1=0,
 \qquad H^2+r\rho H-rb_2=0.
 \label{eq:uncoupledquad}
\end{equation}
Choosing the positive restoring roots,
\begin{equation}
 A=\frac{\sqrt{r^2\rho^2+4rb_1}-r\rho}{2},\quad
 H=\frac{\sqrt{r^2\rho^2+4rb_2}-r\rho}{2}.
 \label{eq:uncoupledroots}
\end{equation}
The linear equations~\eqref{eq:hlinear} reduce to $(\rho+A/r)p_1=-b_1m_1$ and $(\rho+H/r)q_2=-b_2m_2$, with $q_1=p_2=0$. Because $b_1=A(\rho+A/r)$ and $b_2=H(\rho+H/r)$, it follows that $p_1=-Am_1$, $q_2=-Hm_2$. The controls thus become
\begin{equation}
 u_1^*=-\frac Ar(x_1-m_1),\qquad
 u_2^*=-\frac Hr(x_2-m_2),
 \label{eq:uncoupledpol}
\end{equation}
with diagonal $B=r^{-1}\operatorname{diag}(A,H)$ and $\sigma=0$. This is the expected independent OU limit.

\subsection{Exchange-symmetric reciprocal coordination: \texorpdfstring{$b_1=b_2$}{b1=b2}}\label{app:symmetry}
For $s_1=s_2=+1$ and $b_1=b_2=b$, define $S=\bigl(\begin{smallmatrix}0&1\\1&0\end{smallmatrix}\bigr)$. Then $Q_2=SQ_1S$ and $E_2=SE_1S$, so the Riccati equations are invariant under $P_1\mapsto SP_2S$, $P_2\mapsto SP_1S$. The exchange-symmetric branch satisfies
\begin{equation}
 P_2=SP_1S=\begin{pmatrix}E&C\\C&A\end{pmatrix},
 \qquad F=E,\quad G=C,\quad H=A.
 \label{eq:exchangesym}
\end{equation}
It follows that $B=r^{-1}\bigl(\begin{smallmatrix}A&C\\C&A\end{smallmatrix}\bigr)$ is symmetric and $\sigma=0$. This statement concerns the symmetric branch: exchange symmetry alone does not establish global Nash-branch uniqueness.

\subsection{Short-foresight limit of Nash feedback}\label{app:myopic}
For either coupling class and large $\rho$, the Riccati system has $P_i=Q_i/\rho+O(\rho^{-3})$. Equation~\eqref{eq:hlinear} similarly gives $\bm h_i=\bm d_i/\rho+O(\rho^{-3})$. Hence
\begin{equation}
 u_i^*=-\frac{1}{r\rho}\bm e_i^T(Q_i\bm x+\bm d_i)
 +O(\rho^{-3})=-\frac1{r\rho}\frac{\partial c_i}{\partial x_i}+O(\rho^{-3}).
 \label{eq:myopic}
\end{equation}
In particular, $B=O(\tau)$, with relaxation rates of order $a\tau/r$ and $b_i\tau/r$. For reciprocal coordination, $\partial_{x_i}c_i=\partial_{x_i}U$, with $U$ defined in Eq.~\eqref{eq:potential}; thus the leading drift is $-(r\rho)^{-1}\nabla U$. For conformist--contrarian coupling, the leading drift matrix $B$ already has unequal off-diagonal entries, $B_{12}=-a/(r\rho)$ and $B_{21}=+a/(r\rho)$, up to corrections of order $O(\rho^{-3})$. This leading response through gradients of individual state costs is analogous to the ``selfish energies'' viewpoint for nonreciprocal dynamics \citelink{loosjack2026}. The strictly infinite-$\rho$ process has vanishing drift and no normalizable stationary distribution; all entropy-production limits refer to finite $\rho$ followed by the asymptotic limit.

\subsection{Preferred-state dependence of affine feedback}\label{app:affine}
Using $\bm h_1=(p_1,q_1)^T$, $\bm h_2=(p_2,q_2)^T$, Eq.~\eqref{eq:hlinear} yields four linear equations, valid for either sign choice:
\begin{align}
 \rho p_1&=-b_1m_1-\frac{Ap_1+Cq_2+Gq_1}{r},\nonumber\\
 \rho q_1&=-\frac{Cp_1+Eq_2+Hq_1}{r},\nonumber\\
 \rho p_2&=-\frac{Gq_2+Fp_1+Ap_2}{r},\nonumber\\
 \rho q_2&=-b_2m_2-\frac{Hq_2+Gp_1+Cp_2}{r}.
 \label{eq:affine4}
\end{align}
Solving this linear system gives the affine drift term $\bm f_0=-(p_1,q_2)^T/r$. All terms containing $m_i$ affect $\bm f_0$ but never $B$; centering around $\bm\mu=B^{-1}\bm f_0$ produces Eq.~\eqref{eq:centered}. Thus the stationary covariance, entropy production, and information flow are independent of the fixed anchor locations within the stated linear model.

\section{Cooperative-planner comparison}\label{app:planner}
A separate benchmark keeps the same original dynamics and costs but allows a single planner to minimize the total expected discounted cost $J_1+J_2$. The total running cost is
\begin{equation}
 L_{\mathrm{coop}}=\tfrac12\bm x^TQ_{\mathrm{coop}}\bm x
 +\bm d_{\mathrm{coop}}^T\bm x+c_{\mathrm{coop}}^{(0)}
 +\tfrac r2\bm u^T\bm u,
 \quad Q_{\mathrm{coop}}=Q_1+Q_2.
 \label{eq:coopcost}
\end{equation}
Here $\bm d_{\mathrm{coop}}=\bm d_1+\bm d_2$ and $c_{\mathrm{coop}}^{(0)}=c_1^{(0)}+c_2^{(0)}$. With $V_{\mathrm{coop}}=\tfrac12\bm x^TP_{\mathrm{coop}}\bm x+\bm h_{\mathrm{coop}}^T\bm x+k_{\mathrm{coop}}$, $P_{\mathrm{coop}}=P_{\mathrm{coop}}^T$, the HJB first-order condition gives $\bm u^*_{\mathrm{coop}}=-(P_{\mathrm{coop}}\bm x+\bm h_{\mathrm{coop}})/r$. Matching coefficients yields
\begin{equation}
 \rho P_{\mathrm{coop}}=Q_{\mathrm{coop}}-\frac{P_{\mathrm{coop}}^2}{r},\qquad
 \rho\bm h_{\mathrm{coop}}=\bm d_{\mathrm{coop}}-\frac{P_{\mathrm{coop}}\bm h_{\mathrm{coop}}}{r}.
 \label{eq:coopric}
\end{equation}
For the stabilizing solution, $B_{\mathrm{coop}}=P_{\mathrm{coop}}/r$ is symmetric, so the OU stationary probability current vanishes for isotropic diffusion and $\sigma_{\mathrm{coop}}=0$. In reciprocal coordination, the cooperative planner retains the interaction between the states but adjusts them through a common optimization problem; its objective differs from the instantaneous potential $U$ in Eq.~\eqref{eq:potential}, since summing the individual costs counts the shared coordination penalty twice. In conformist--contrarian coupling, the opposing interaction penalties satisfy
\begin{equation}
 \frac{a}{2}(x_1-x_2)^2+\frac{a}{2}(x_1+x_2)^2
 =a(x_1^2+x_2^2).
 \label{eq:coopcancel}
\end{equation}
The cross terms therefore cancel in the cooperative objective, making $Q_{\mathrm{coop}}$ diagonal and the resulting optimal feedback decoupled. In both coupling classes, the cooperative benchmark retains the original dynamics and individual costs but replaces separate Nash optimization with joint optimization.

\section{Numerical method and parameter choices}\label{app:numerical}
Equations~\eqref{eq:six} were solved numerically using \texttt{scipy.optimize.root}, with continuation in increasing $\tau$ from the short-foresight approximation $P_i\simeq Q_i/\rho$. A solution was accepted only if its Riccati residual was below $2\times10^{-8}$ and all eigenvalues of $B$ had positive real part; independent single-agent best-response calculations, using \texttt{scipy.linalg.solve\_continuous\_are} with the opponent's equilibrium feedback held fixed, were performed at sampled points along each scan. Stationary covariances were obtained using \texttt{scipy.linalg.solve\_continuous\_lyapunov}, entropy production was evaluated from Eqs.~\eqref{eq:ep}--\eqref{eq:epnormalized}, and the information flow from Eq.~\eqref{eq:infoflow}. For reciprocal coordination, the short-foresight values of $\Delt$ in \figrefs{fig:coordscaling}{fig:mechanism} were evaluated from the exact identity~\eqref{eq:identity}, which avoids the cancellation in the direct difference $(C-G)/r$; the two evaluations agree to within $2\times10^{-7}$. The same identity was used for $\Iflow_1$ in \figref{fig:mechanism}(b). Numerical calculations and figures were produced using Python, NumPy, SciPy, and Matplotlib.

Unless specified otherwise, we use $a=r=D=1$ and $m_1=m_2=0$. For reciprocal coordination, the main parameter choice is $b_1=0.1$, $b_2=1$, while for conformist--contrarian coupling it is $b_1=b_2=0.5$. The main calculations use 240 logarithmically spaced planning timescales over $0.006\leq\tau\leq10^4$. The short-foresight exponents were fitted over $0.009\lesssim\tau\lesssim0.028$ for reciprocal coordination and $0.009\lesssim\tau\lesssim0.035$ for conformist--contrarian coupling. The self-anchoring scans vary $b_2$ at fixed $b_1=0.1$, with the irreversibility landscape covering $0.1\leq b_2\leq2.5$.

For \figref{fig:mechanism}(a), the modified game adds $g_1(x_2-n_1)^2/2$ with $g_1=0.5$ to the cost of agent 1. This changes only $(Q_1)_{22}$ from $a$ to $a+g_1$ (the reference value $n_1$ enters only the affine terms), so Eq.~\eqref{eq:identity} still applies. It was solved with the same procedure and checks on the points of the main grid with $\tau\leq0.1$, and fitted over the same short-foresight interval.

The computed solutions satisfied the Riccati equations to numerical precision and remained stable throughout the sampled parameter ranges. Independent Nash best-response checks agreed at the parameter points where they were performed.

\end{document}